\documentclass[12pt]{article}
\usepackage{textcomp}

\usepackage[ngerman,english]{babel}

\usepackage{makeidx}
\usepackage{booktabs}
\usepackage{makeidx}

\usepackage{algorithm}
\usepackage{algorithmic}
\usepackage{listings}

\usepackage{comment}
\usepackage{amsmath}
\usepackage{amsfonts}
\usepackage{amssymb}
\usepackage{amsthm}
\usepackage{epsf}
\usepackage{psfrag}
\usepackage{graphicx}
\usepackage[subfigure]{caption2}
\usepackage{subfigure}
\usepackage{palatino}
\usepackage{rotating}
\usepackage{wrapfig}

\usepackage{ziffer}

\usepackage{ngerman}
\usepackage{t1enc}
\usepackage[ansinew]{inputenc}

\usepackage{hyperref}

\newcommand{\R}{\mathbb{R}}

\usepackage{color}
\definecolor{hellgrau}{gray}{0.96}
\definecolor{darkblue}{rgb}{0,0,.7}
\definecolor{darkred}{rgb}{.6,0,0}
\definecolor{darkgreen}{rgb}{.30,.69,.10}
\definecolor{orange}{rgb}{0.8,0.3,0.3}

\title{Prospects and limits of SIR-type Mathematical Models to Capture the COVID-19 Pandemic}
\author{G\"unter B\"arwolff\footnote{mailto:baerwolf@math.tu-berlin.de}\\
	Technische Universit\"at Berlin}

\date{}

\begin{document}

\maketitle

\begin{abstract}
For the description of a pandemic mathematical models could be interesting. Both for physicians and
politicians as a base for decisions to treat the disease. 
The responsible estimation of parameters is a main issue of mathematical pandemic models.
Especially a good choice of $\beta$
as the number of others that one infected person
encounters per unit time (per day) influences the adequateness of the results of the model.
For the actual COVID-19 pandemic some aspects of the
parameter choice will be discussed.
Because of the incompatibility of the data of the Johns-Hopkins-University \cite{jhu} to the data
of the German Robert-Koch-Institut we use the COVID-19 data of the 
European Centre for Disease Prevention and Control \cite{ecdc} (ECDC) as a base for the parameter estimation.
Two different mathematical methods for the data analysis will be discussed in this paper and possible
sources of trouble will be shown.

As example of the parameter choice serve the data of the USA and the UK.
The resulting parameters will be used 
estimated and used 
in W.\,O. Kermack and A.\,G. McKendrick's SIR model\cite{kak}.  
Strategies for the commencing and ending of
social and economic shutdown measures are discussed.

The numerical solution of the ordinary differential equation system of the
modified SIR model is being done with a Runge-Kutta integration method of fourth order \cite{gb}.

At the end the applicability of the SIR model could be shown essentially.
Suggestions about appropriate points in time at which to commence with
lockdown measures based on the acceleration rate of infections conclude the paper.
This paper is an improved sequel of \cite{gb1}.

\end{abstract}

\section{The mathematical SIR model}

Let us recollect something about the the model. $I$ denotes the infected people, $S$ stands for
the susceptible and $R$ denotes the recovered people.
The dynamics of infections and recoveries can be approximated by the ODE system
\begin{eqnarray} \label{eq1}
  \frac{dS}{dt} & = & -\beta \frac{S}{N} I  \\ \label{eq2}
  \frac{dI}{dt} & = & \beta \frac{S}{N} I - \gamma I \\ \label{eq3}
  \frac{dR}{dt} & = & \gamma I\;.
\end{eqnarray}
We understand $\beta$
as the number of others that one infected person
encounters per unit time (per day). $\gamma$ is the reciprocal value of
the typical time from infection to recovery.
$N$ is the total number of people involved in the epidemic disease and there is
$N = S + I + R$. 
The evenly distribution of members of the species $S$, $I$ and $R$ is an important 
assumption for the SIR model\footnote{This is usually not given in reality.}.

The empirical data currently available suggests that the corona infection typically lasts for some 14 days.
This means $\gamma = 1/14 \approx 0,07$.

The choice of $\beta$ is more complicated and will be considered in the next section.
It should be noted, that there are a lot of modifications of the SIR model adding other values then
$I$, $S$ or $R$, but the main behavior of the model will be the same.

\section{The estimation of $\beta$ based on real data}
We use the 
European Centre for Disease Prevention and Control \cite{ecdc} as a data for
the COVID-19 infected people for the period from December 31st 2019 to April 8th 2020.

At the beginning of the pandemic the quotient $S/N$ is nearly equal to 1.
Also, at the early stage no-one has yet recovered.
Thus we can describe the early regime by the equation
\[ \frac{d I}{dt} = \beta I   \]
with the solution
\begin{equation}\label{eqexp}
  I(t) = I_0 \exp(\beta t) \;.
\end{equation}

We are looking for periods in the spreadsheets of infected people per day where the
course can be described by a function of type \eqref{eqexp}. Starting with a spreadsheet
like

\bigskip\noindent
\begin{center}
%\begin{tabular}{|l|l|} \toprule
\begin{tabular}{|l|l|} \hline
%day & number of infected people\\ \toprule
day & number of infected people\\ \hline\hline
$t_1$ & $I_1$ \\
$t_2$ & $I_2$ \\
\vdots & \vdots  \\
%$t_k$ & $I_k$ \\ \bottomrule
$t_k$ & $I_k$ \\ \hline
\end{tabular}
\end{center}
\bigskip\noindent

for a certain country and a chosen period $[t_1,t_k]$
with my favored method 
We search for the minimum of the functional
\[     F(I_0, \beta) = \sum_{j=1}^k [I_0 \exp(\beta t_j) - I_j]^2  \;,\]
i.e.
\begin{equation}\label{nlopt}
     \min_{(I_0,\beta)\in \R^2} F(I_0, \beta)   \;.
\end{equation}
We solved this non-linear minimum problem with the damped Gauss-Newton method (see \cite{gb}). After some
numerical tests we found the subsequent results for the considered countries.
Thereby we chose different periods for the countries with the aim to approximate
the infection course in a good quality. The following figures show the graphs and the
evaluated parameter of the USA and the UK.

\begin{figure}[h]
\begin{minipage}[t]{0.49\textwidth}
\includegraphics[width=8.7cm]{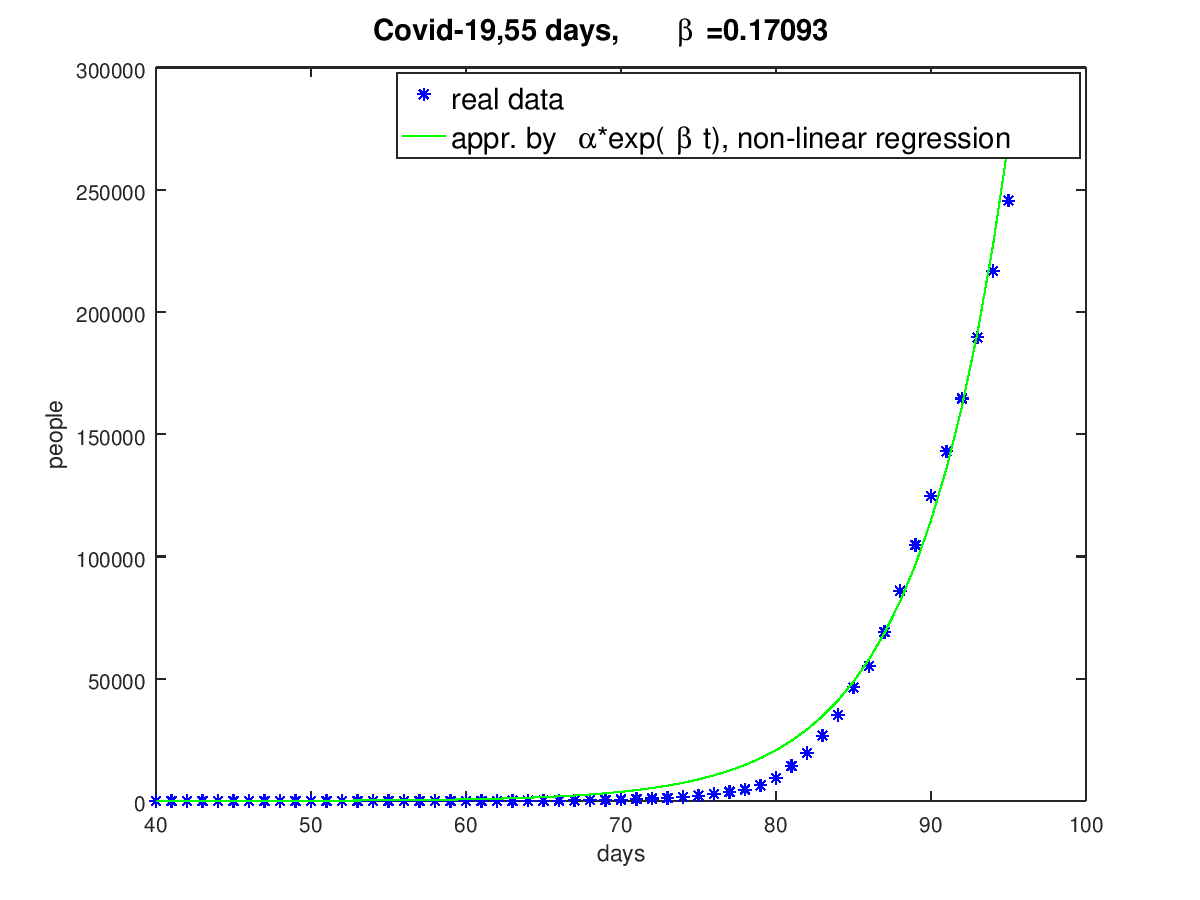}
\caption{\label{fig1} USA course from February 10th 2020 to April 4th 2020}
\end{minipage}
\hfill
\begin{minipage}[t]{0.49\textwidth}
\includegraphics[width=8.7cm]{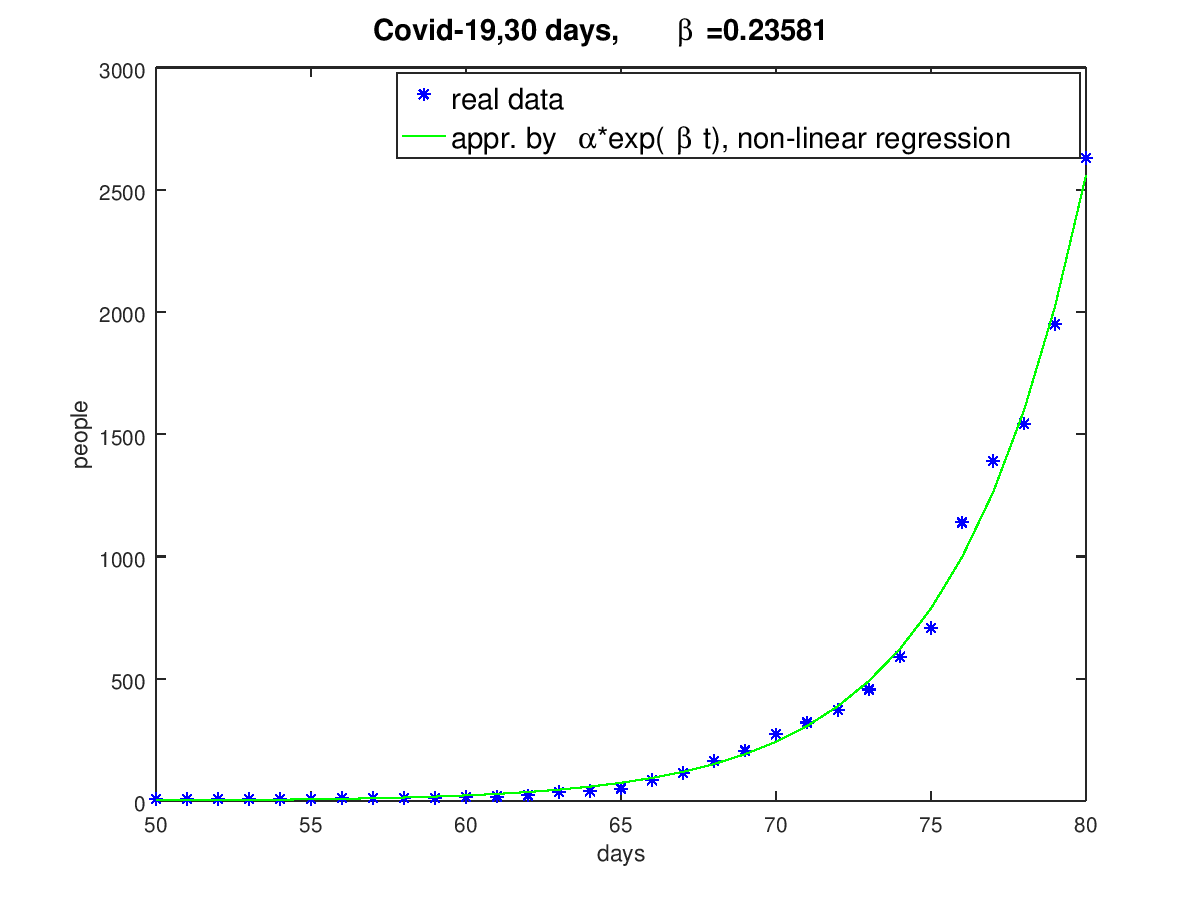}
\caption{\label{fig2} UK course from February 20th 2020 to March 20th 2020}
\end{minipage}
\end{figure}

It must be said that evaluated $\beta$-values are related to the stated period.
For the iterative Gauss-Newton method we guessed the respective periods for every country
by a visual inspection of the graphs of the infected people over days.

Especially in medicine, psychology and other life sciences the logarithm behavior of data
was readily considered. 

Instead of the above table of values the following logarithmic one was used.

\bigskip\noindent
\begin{center}
%\begin{tabular}{|l|l|} \toprule
\begin{tabular}{|l|l|} \hline
%day & number of infected people\\ \toprule
day & log(number of infected people)\\ \hline
$t_1$ & $\log I_1$ \\
$t_2$ & $\log I_2$ \\
\vdots & \vdots  \\
%$t_k$ & $\log I_k$ \\ \bottomrule
$t_k$ & $\log I_k$ \\ \hline
\end{tabular}
\end{center}
\bigskip\noindent

The logarithm of \eqref{eqexp} leads to
\[   \log I(t) = \log I_0 + \beta t  \]
and based on the logarithmic table
the functional
\[     L(I_0, \beta) = \sum_{j=1}^k [\log I_0  + \beta t_j - \log I_j]^2  \;,\]
is to minimize. The solution of this linear optimization problem is trivial and it is available
in most of computer algebra systems as a ''block box'' of the logarithmic-linear regression.

The following figures show the results for the same periods as above for the USA and the UK.

\begin{figure}[h]
\begin{minipage}[t]{0.49\textwidth}
\includegraphics[width=8.7cm]{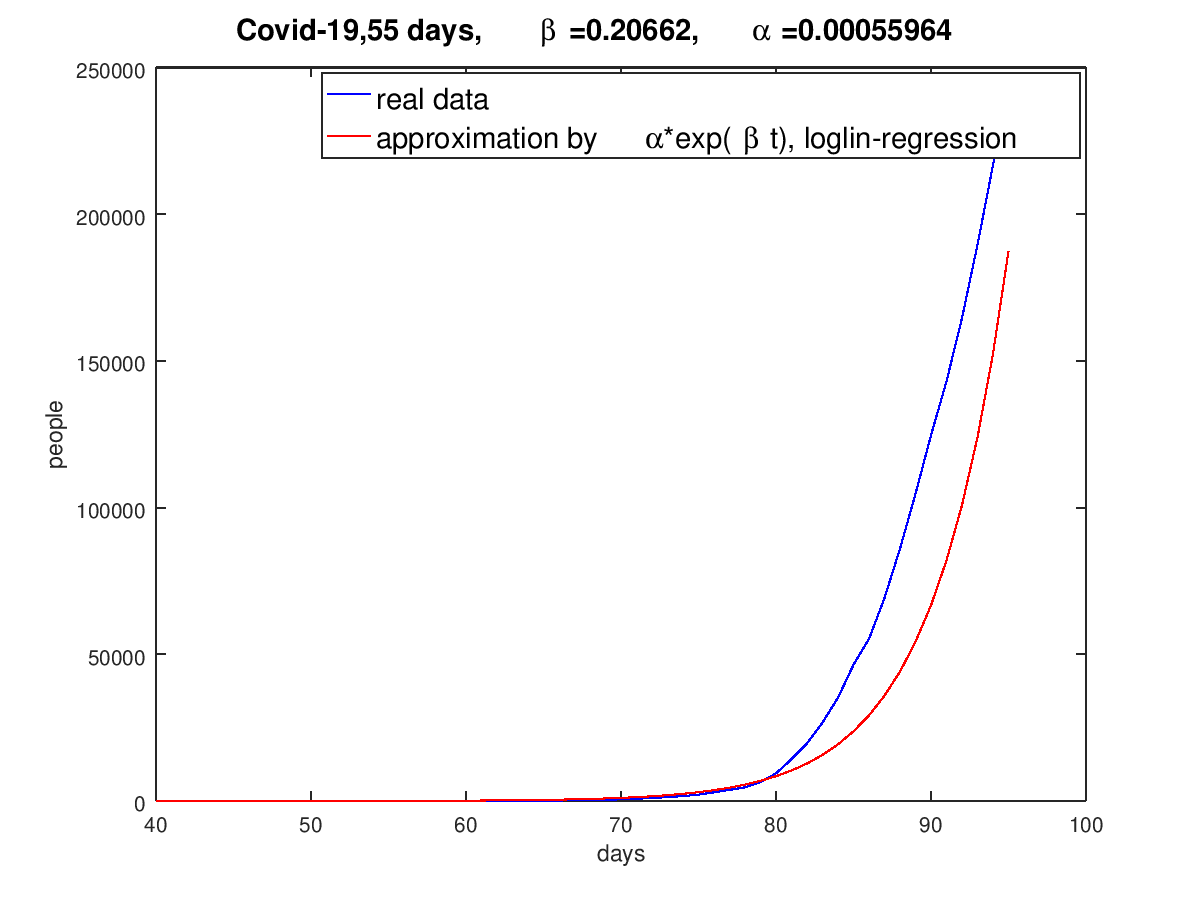}
\caption{\label{fig11} log-lin-result of the USA (February 10th 2020 to April 4th 2020)}
\end{minipage}
\hfill
\begin{minipage}[t]{0.49\textwidth}
\includegraphics[width=8.7cm]{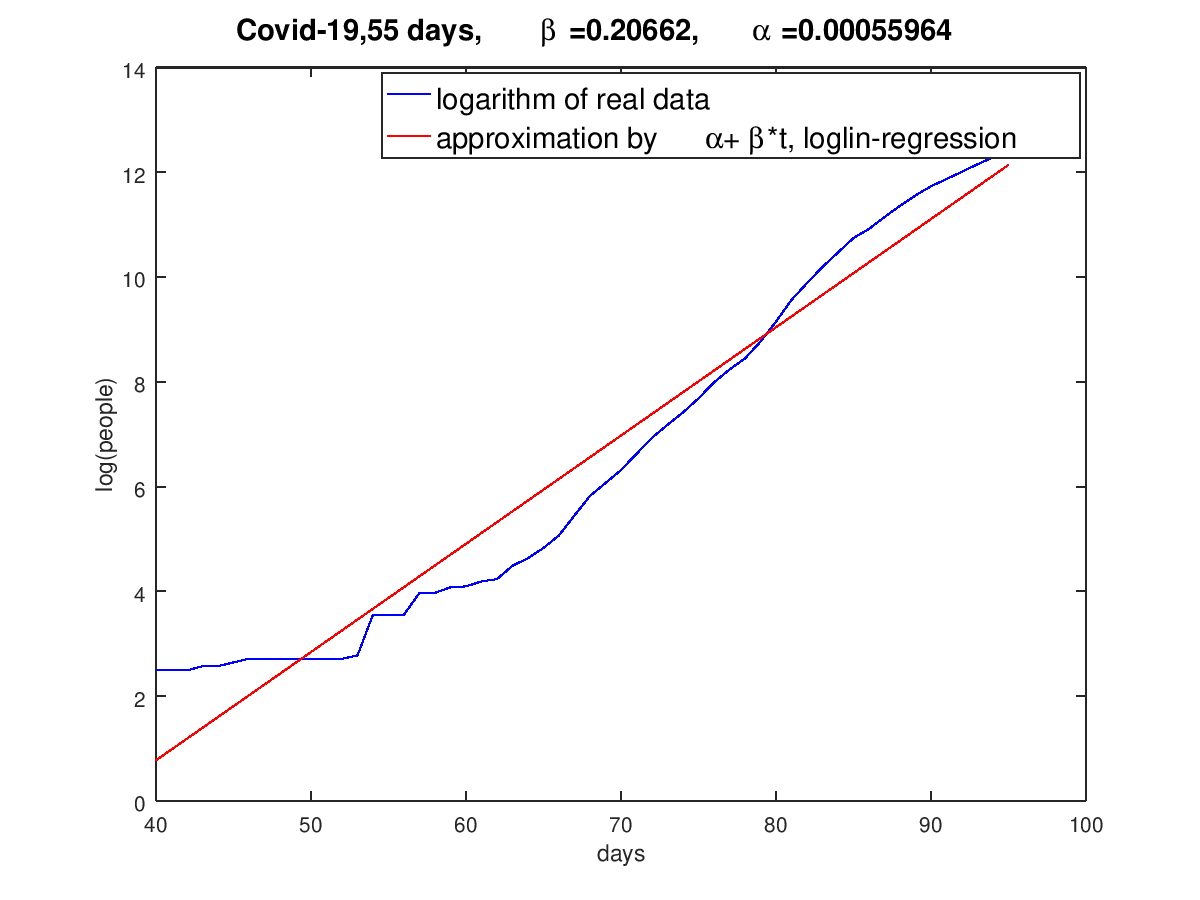}
\caption{\label{fig12} Logarithm of the USA result (February 10th 2020 to April 4th 2020)}
\end{minipage}
\end{figure}

\begin{figure}[h]
\begin{minipage}[t]{0.49\textwidth}
\includegraphics[width=8.7cm]{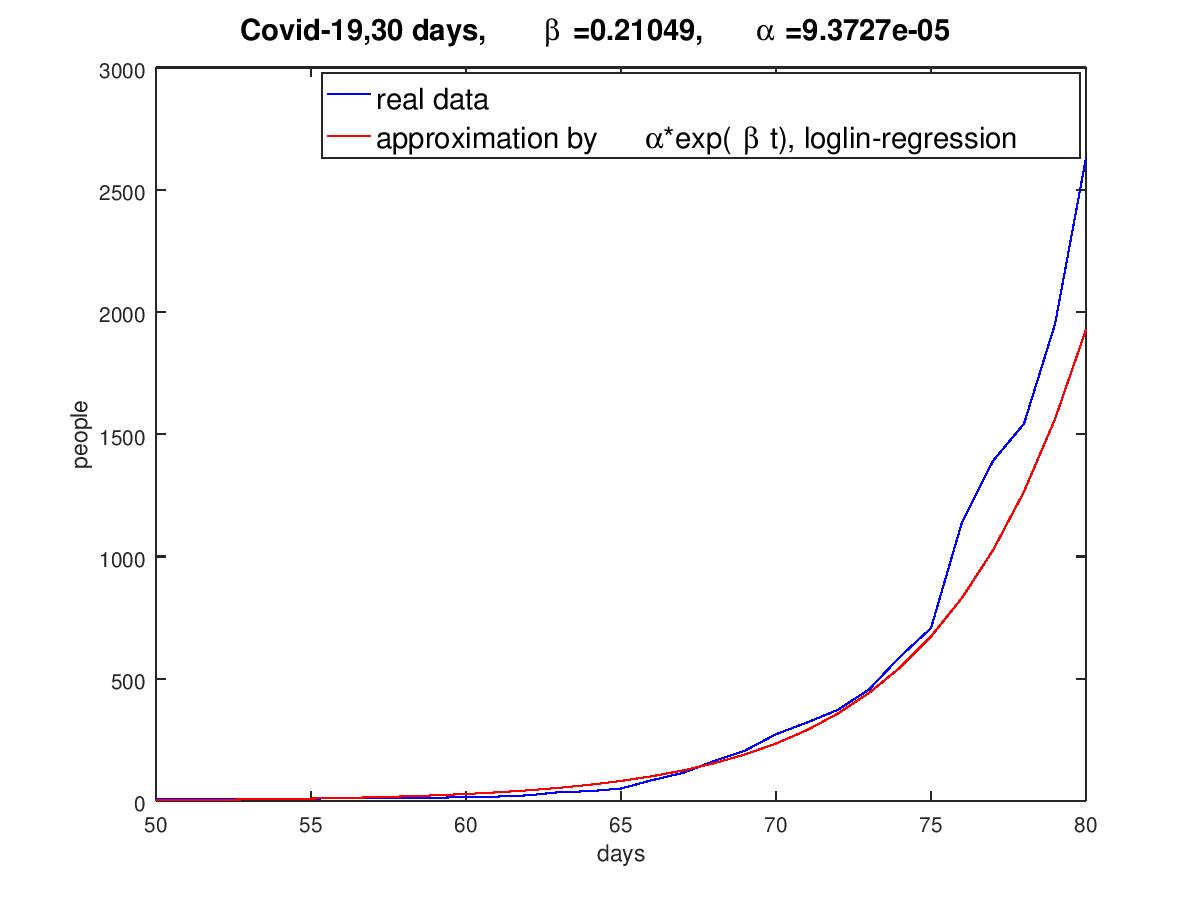}
\caption{\label{fig9} log-lin-result of the UK (February 20th 2020 to March 20th 2020)}
\end{minipage}
\hfill
\begin{minipage}[t]{0.49\textwidth}
\includegraphics[width=8.7cm]{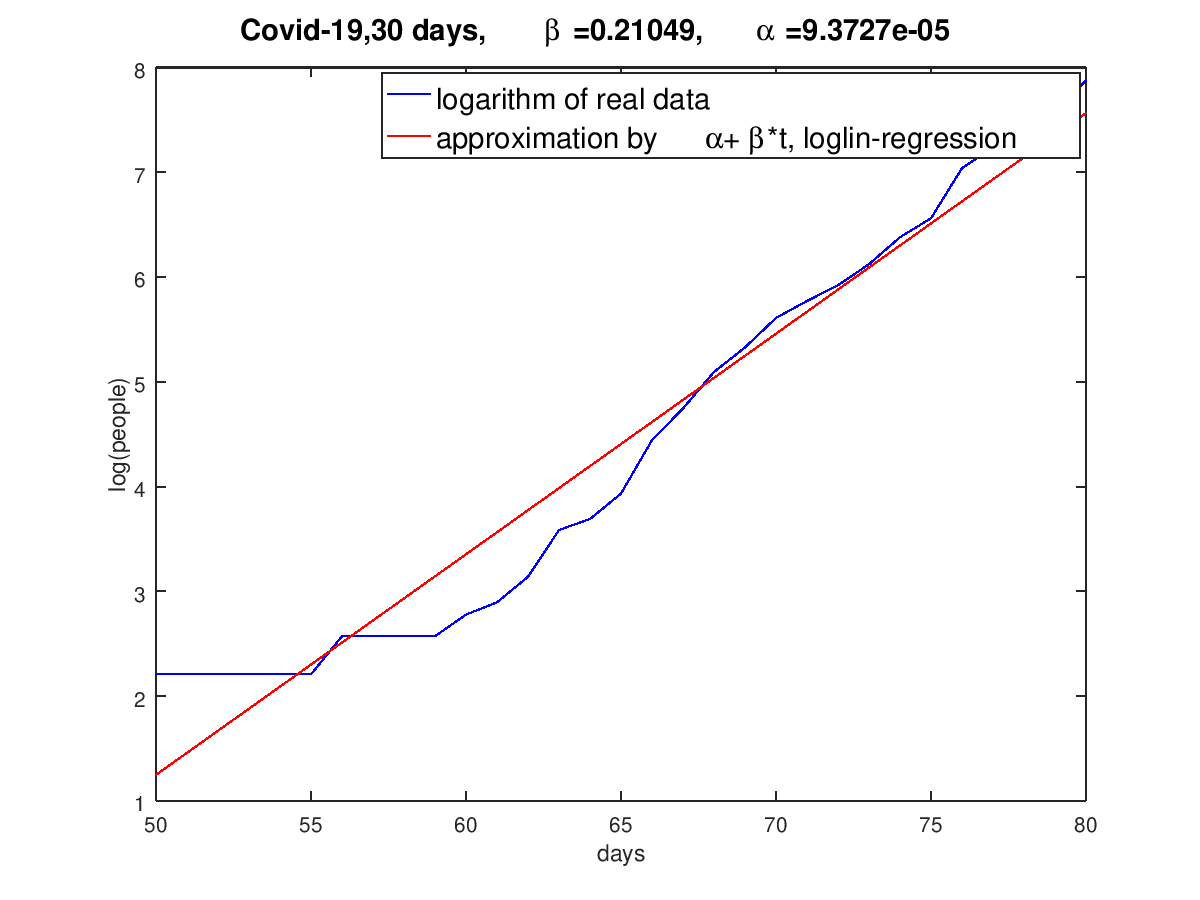}
\caption{\label{fig10} Logarithm of the UK result (February 20th 2020 to March 20th 2020)}
\end{minipage}
\end{figure}

Figures \ref{fig11}-\ref{fig10}
show that the logarithmic-linear regression implies poor results. Thus, the non-linear
optimization problem \eqref{nlopt} is to choose as the favored method for the estimation
of $I_0$ and $\beta$. 

\begin{figure}[h]
\begin{minipage}[t]{0.49\textwidth}
\includegraphics[width=8.5cm]{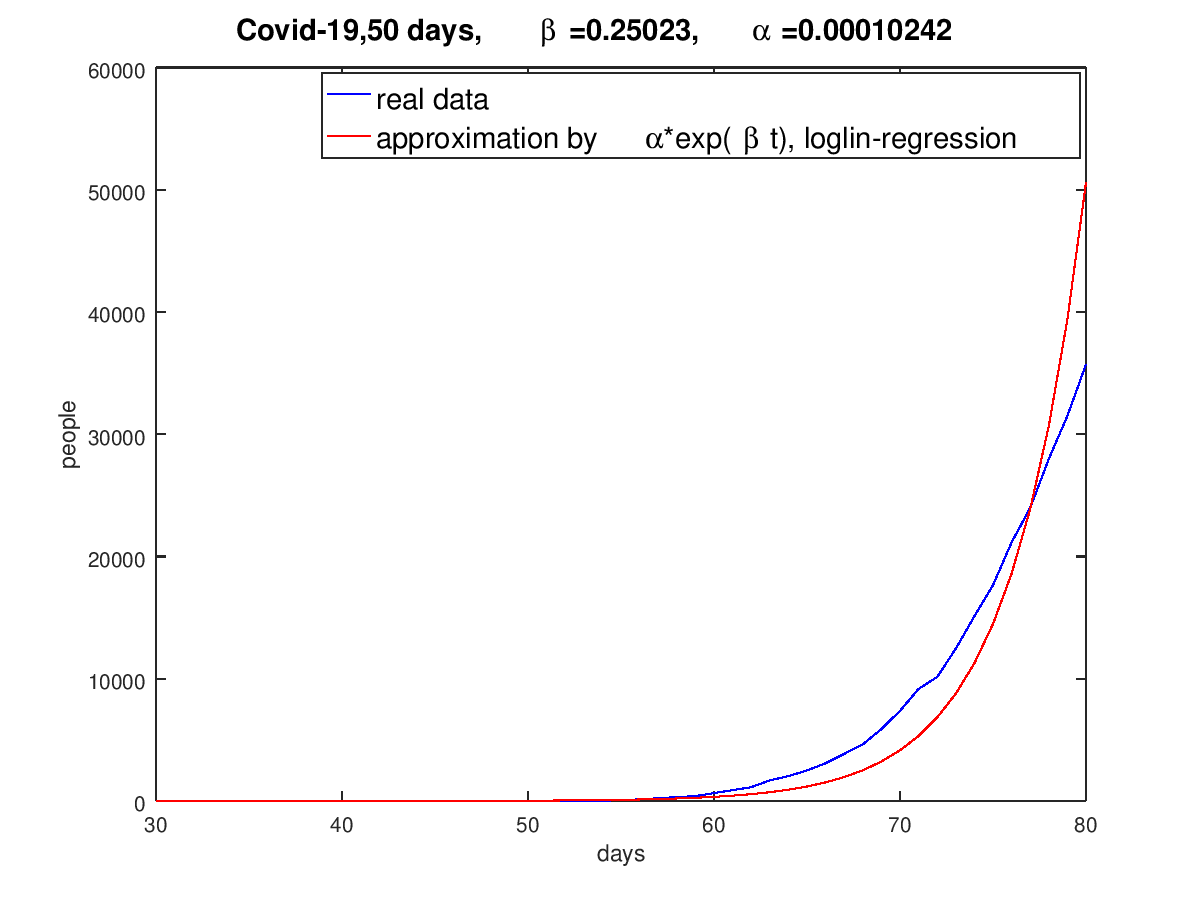}
\caption{\label{fig13} log-lin-result of Italy (January 31st 2020 to March 20th 2020)}
\end{minipage}
\hfill
\begin{minipage}[t]{0.49\textwidth}
\includegraphics[width=8.5cm]{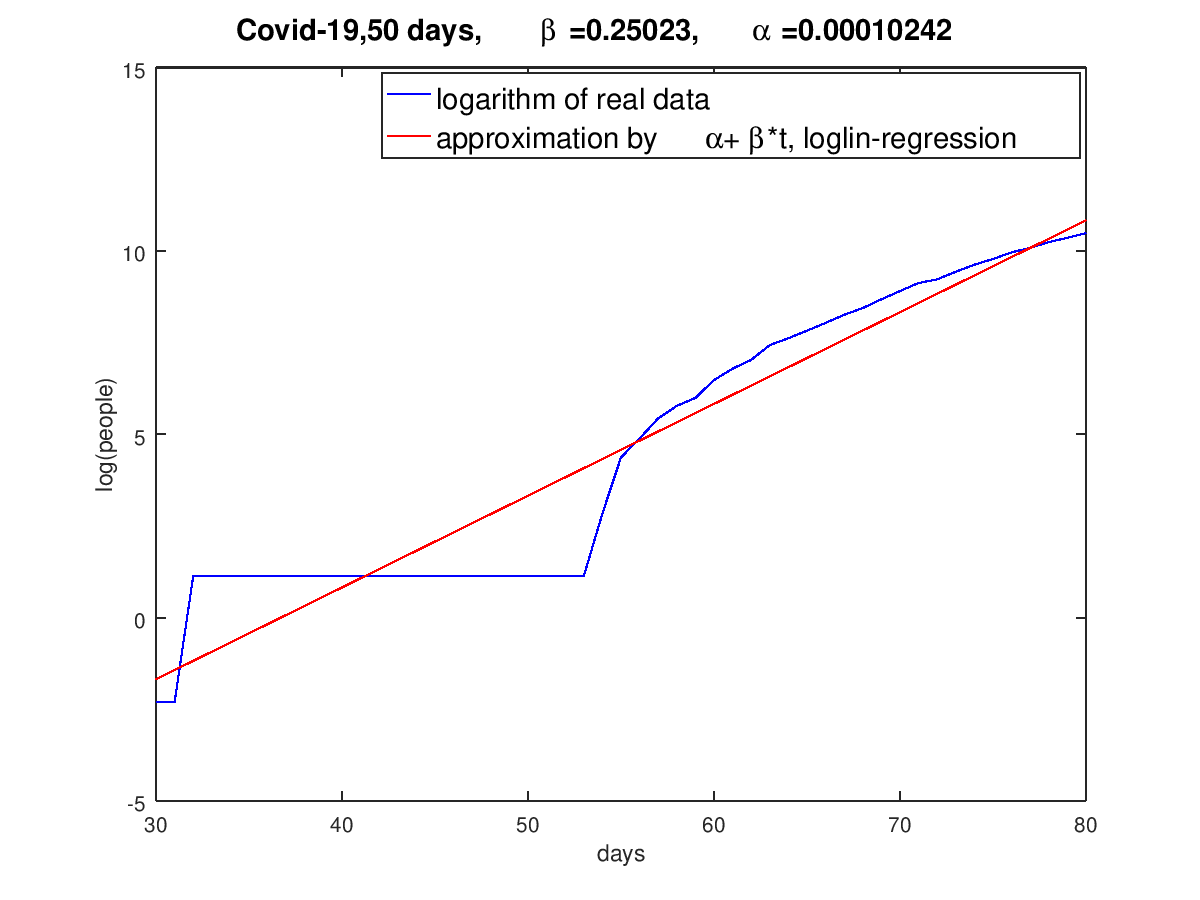}
\caption{\label{fig14} Logarithm of the Italian result (January 31st 2020 to March 20th 2020)}
\end{minipage}
\end{figure}

We found some notes on the parameters of Italy in the literature,
for example $\beta = 0.25$, and we are afraid that this is a result of the logarithmic-linear regression.
Our result for Italy is pictured in fig. \ref{fig13} and fig. \ref{fig14}.

\section{Some numerical computations for the USA and the UK}

\begin{figure}[h]
\begin {center}
\includegraphics[width=10cm]{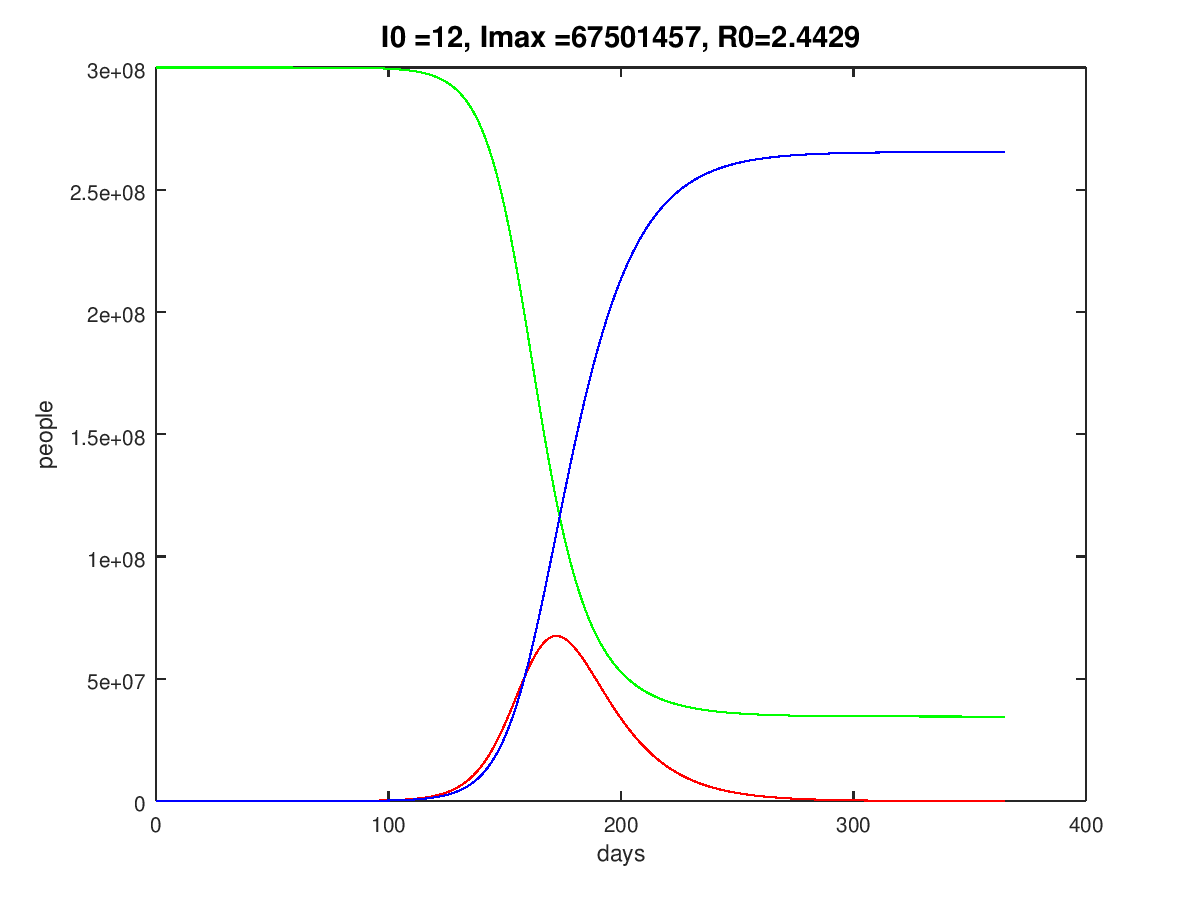}
\caption{\label{figusa1} USA course of one year, starting February 10th 2020,
$S$-green, $I$-red, $R$-blue}
\end {center}
\end{figure}

With the choice of $\beta$-value $0,171$ (see fig. \ref{fig1}) which was evaluated on the basis of the real data
of ECDC and $\gamma = 0,07$ one gets the course of the pandemic dynamics
pictured in fig. \ref{figusa1}.\footnote{$I0$ denotes the initial value
of the $I$ species, that is January 31th 2020. $Imax$ stands
for the maximum of $I$. The total number $N$ for the USA is guessed to be 300 millions.}.
$R_0$ is the basis
reproduction number of persons, infected by the transmission of a pathogen from one infected
person during the infectious time ($R_0 = \beta/\gamma$) in the following figures.

Neither data from ECDC nor the data from the German Robert-Koch-Institut and the data from
the Johns Hopkins University are correct, for we have to reasonably assume that there are
a number of unknown cases. It is guessed that the data covers only 15\% of the real cases.
Considering this we get a slightly changed results and in the subsequent computations
we will include estimated number of unknown cases to the initial values of $I$.

\begin{figure}[h]
\begin {center}
\includegraphics[width=10cm]{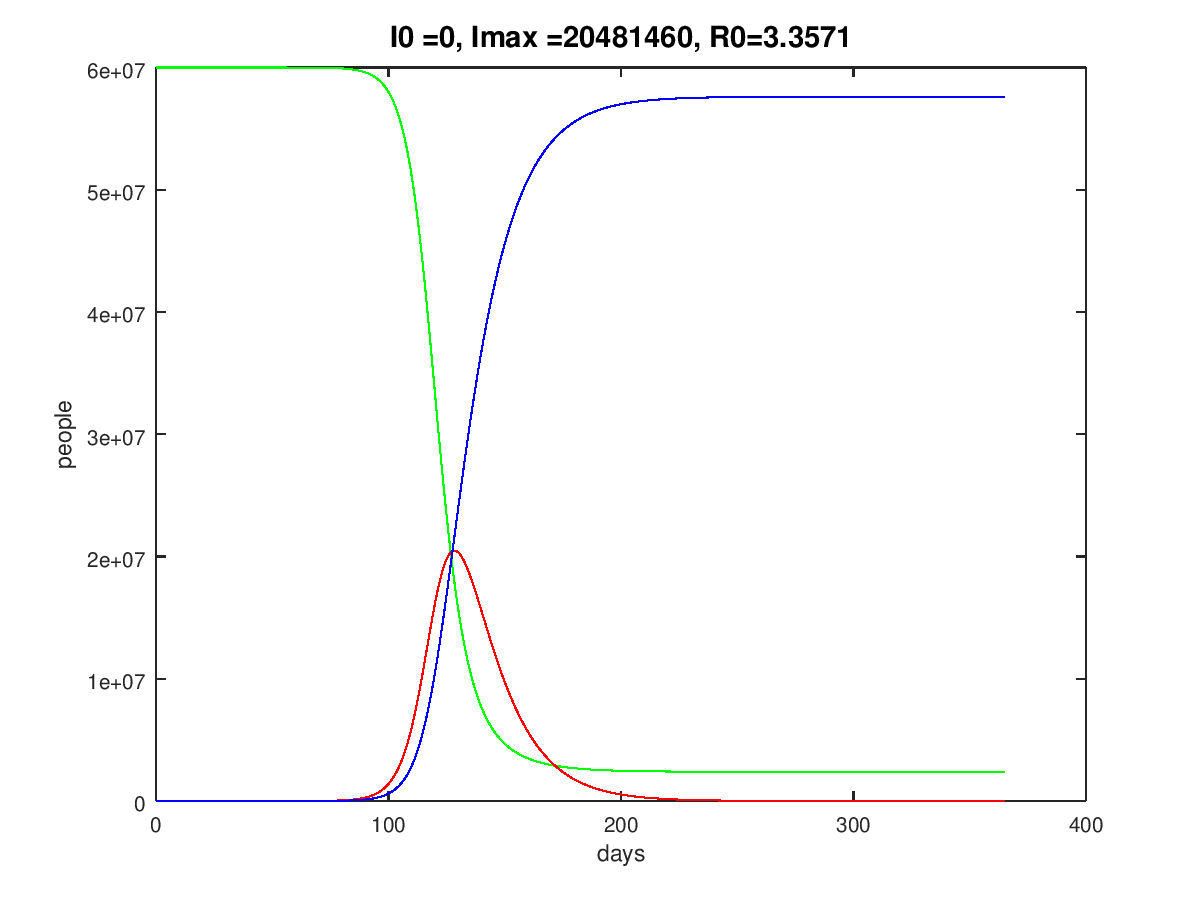}
\caption{\label{figuk1} UK course of one year, starting February 20th 2020,
$S$-green, $I$-red, $R$-blue}
\end {center}
\end{figure}

For the UK we use the  $\beta$-value $0,235$ (see fig. \ref{fig2}) and $\gamma = 0,07$ we get the
course pictured in fig. \ref{figuk1}. $N$ was set to $60$ millions.

\section{Influence of a temporary lockdown and extensive social distancing}

In all countries concerned by the Corona pandemic a lockdown of the
social life is discussed. In Germany the lockdown started at March 16th 2020.
The effects of social distancing to decrease the infection rate can be
modeled by a modification of the SIR model.
The original ODE system \eqref{eq1}-\eqref{eq3}
was modified to
\begin{eqnarray} \label{eq4}
  \frac{dS}{dt} & = & -\kappa\beta \frac{S}{N} I  \\ \label{eq5}
  \frac{dI}{dt} & = & \kappa\beta \frac{S}{N} I - \gamma I \\ \label{eq6}
  \frac{dR}{dt} & = & \gamma I\;.
\end{eqnarray}

$\kappa$ is a function with values in $[0,1]$. For example
\[   \kappa(t) = \left\{\begin{array}{ll} 0,5 & \mbox{for } t_0 \le t \le t_1 \\ 1 & \mbox{for } t > t_1,\; t < t_0
\end{array}\right.  \]
means for example a reduction of the infection rate of 50\% in the period $[t_0,t_1]$ ($\Delta_t = t_1 - t_0$ is the
duration of the temporary lockdown in days).
A good choice of $t_0$ and $t_k$ is going to be complicated.

\begin{figure}[h]
\begin {center}
\includegraphics[width=10cm]{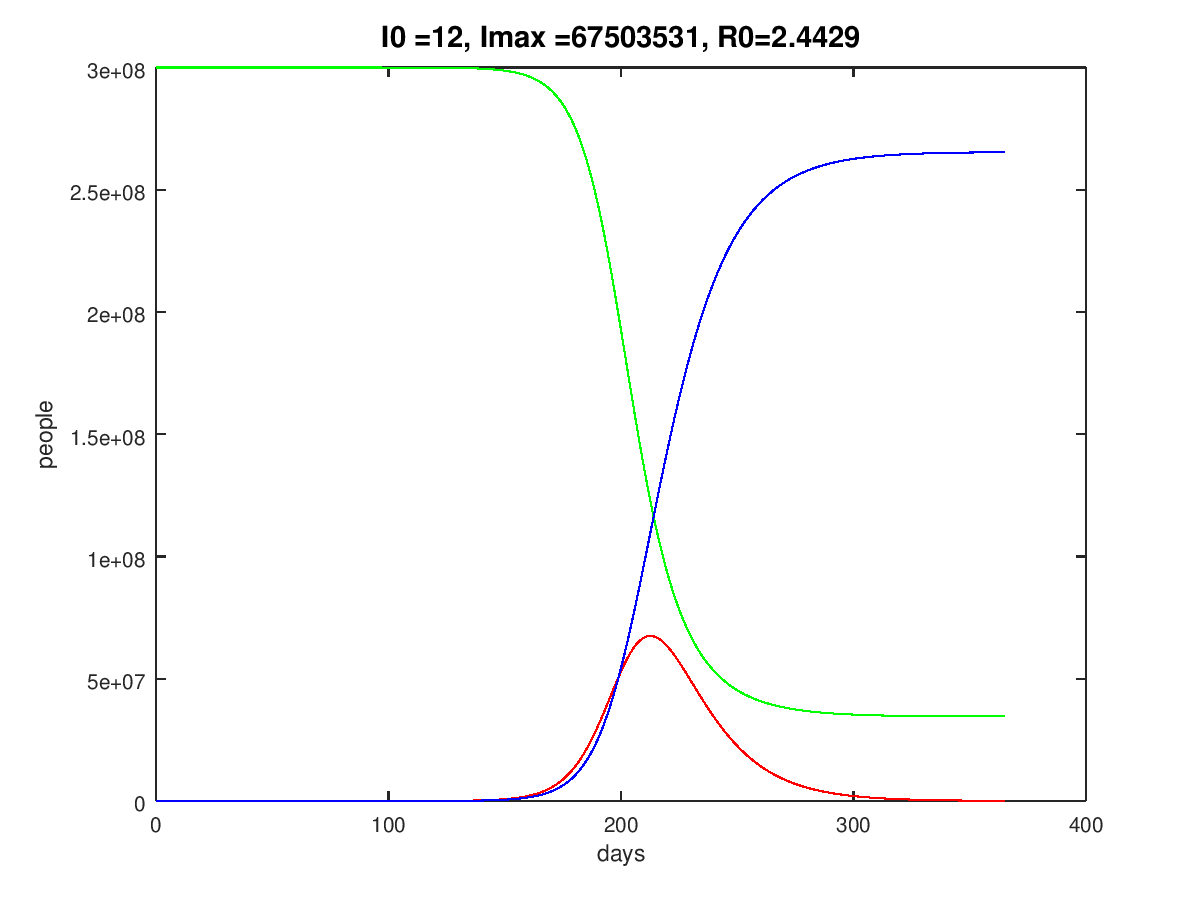}
\caption{\label{figusa2} USA course of one year, starting February 10th 2020,
$S$-green, $I$-red, $R$-blue, 30 days lockdown, starting time March 20th 2020}
\end {center}
\end{figure}

If we respect the chosen starting day of the USA lockdown, March 31st 2020 (this conforms the 50th day of
the concerned year), and we work
with
\[   \kappa(t) = \left\{\begin{array}{ll} 0,2 & \mbox{for } 46 \le t \le 76 \\ 1 & \mbox{for } t > 76,\; t < 46 
\end{array}\right.  \]
we got the result pictured in fig. \ref{figusa2}.

The numerical tests showed that a very early start of the lockdown resulting in a reduction
of the infection rate $\beta$ results in the typical Gaussian curve to be delayed by $I$; however, the
amplitude (maximum value of $I$) doesn't really change.

One knows that development of the infected people looks like a Gaussian curve.
The interesting points in time are those where the acceleration of the numbers of infected
people increases or decreases, respectively.

These are the points in time where the curve of $I$ was changing from a convex to a concave
behavior or vice versa. The convexity or concavity can be controlled by the
second derivative of $I(t)$.

Let us consider equation \eqref{eq2}. By differentiation of \eqref{eq2} and the use
of \eqref{eq1} we get
\begin{eqnarray*}
   \frac{d^2 I}{dt^2} & = & \frac{\beta}{N} \frac{dS}{dt} I + \frac{\beta}{N} S \frac{dI}{dt} - \gamma \frac{dI}{dt} \\
   & = & - \frac{\beta}{N}^2 S I^2 +(\frac{\beta S}{N} -\gamma)(\frac{\beta S}{N} -\gamma)I\\
   & = & [(\frac{\beta S}{N} -\gamma)^2 - (\frac{\beta}{N})^2 S I]I \;.
\end{eqnarray*}
With that the $I$-curve will change from convex to concave if the relation
\begin{equation}\label{k2k}
    (\frac{\beta S}{N} -\gamma)^2 - (\frac{\beta}{N})^2 S I < 0 \Longleftrightarrow
     I > \frac{ (\frac{\beta S}{N} -\gamma)^2 N^2 }{\beta^2 S}
\end{equation}
is valid.
For the switching time follows
\begin{equation}\label{t0f}
     t_0 = \min_t \{t > 0,\, I(t) > (\frac{\beta S(t)}{N} -\gamma)^2 N^2)/(\beta^2 S(t)) \} \;.
\end{equation}
A lockdown starting at $t_0$ (assigning $\beta^* = \kappa\beta$, \;$\kappa \in [0,1[$)
up to a point in time $t_1 = t_0 + \Delta_t$, with $\Delta_t$ as the duration of the
lockdown in days,
will be denoted as a {\bf dynamical lockdown} (for $t> t_1$ $\beta^*$ was reset to the original value $\beta$).

\begin{figure}[h]
\begin {center}
\includegraphics[width=10cm]{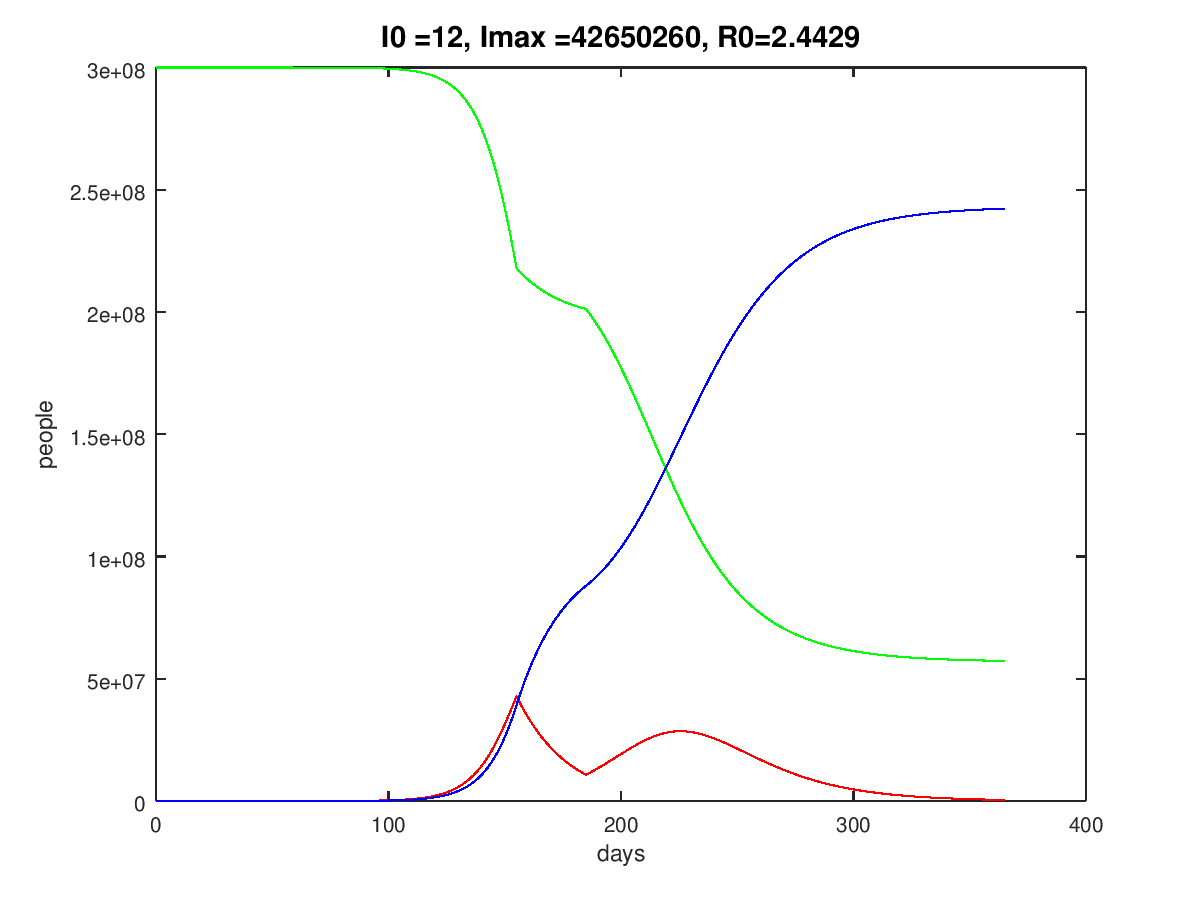}
\caption{\label{figusa3} USA course of one year, starting February 20th 2020, dynamical lockdown, $S$-green, $I$-red, $R$-blue}
\end {center}
\end{figure}

$t_0$ means the point in time up to which the growth rate increases and from which on it decreases.
Fig. \ref{figusa3} shows the result of such a computation of a dynamical lockdown. 
We got $t_0 = 155$ ($\kappa = 0,2$)-
The result is significant. In fig. \ref{figusa4} a typical behavior of $\frac{d^2 I}{dt^2}$
is plotted.

\begin{figure}[h]
\begin {center}
\includegraphics[width=10cm]{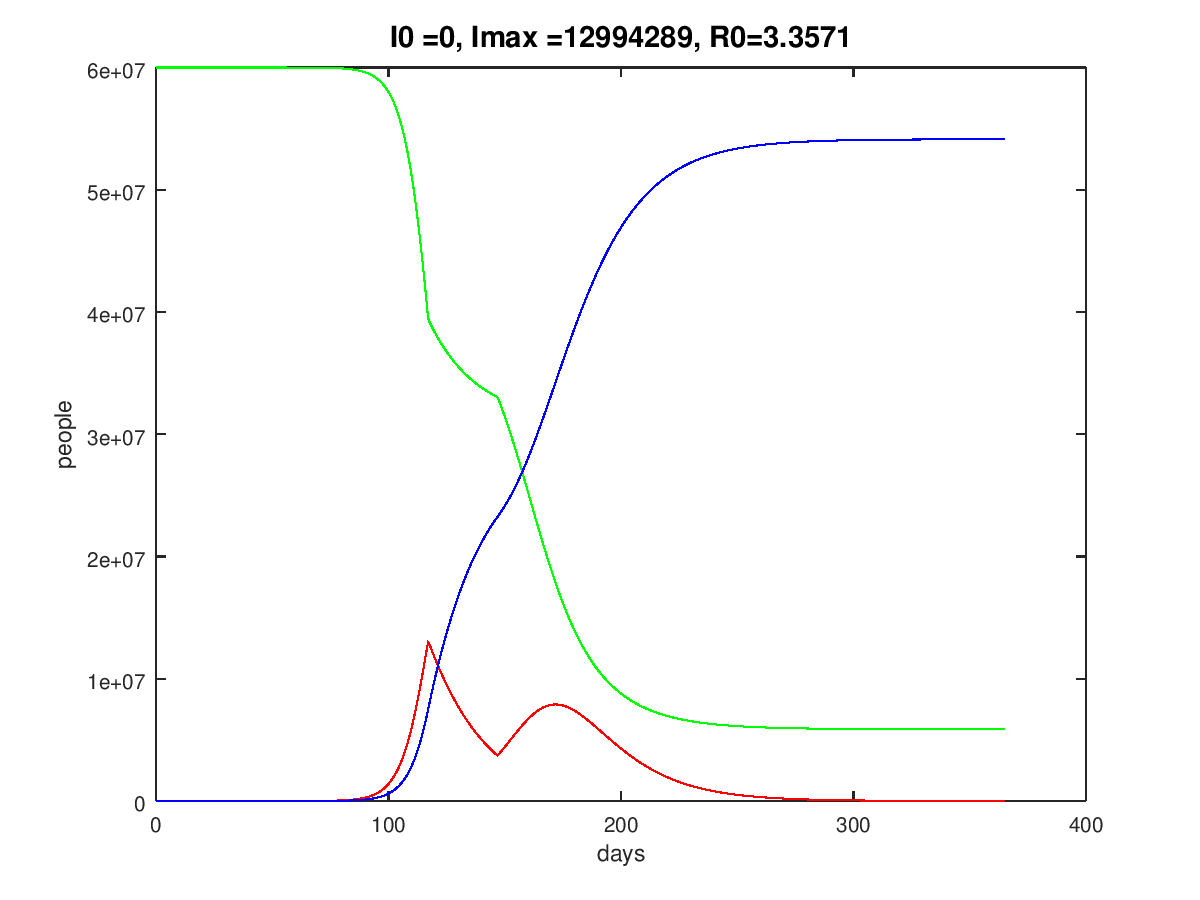}
\caption{\label{figuk2} UK course of one year, starting February 20th 2020, dynamical lockdown, $S$-green, $I$-red, $R$-blue}
\end {center}
\end{figure}

\begin{figure}[h]
\begin {center}
\includegraphics[width=10cm]{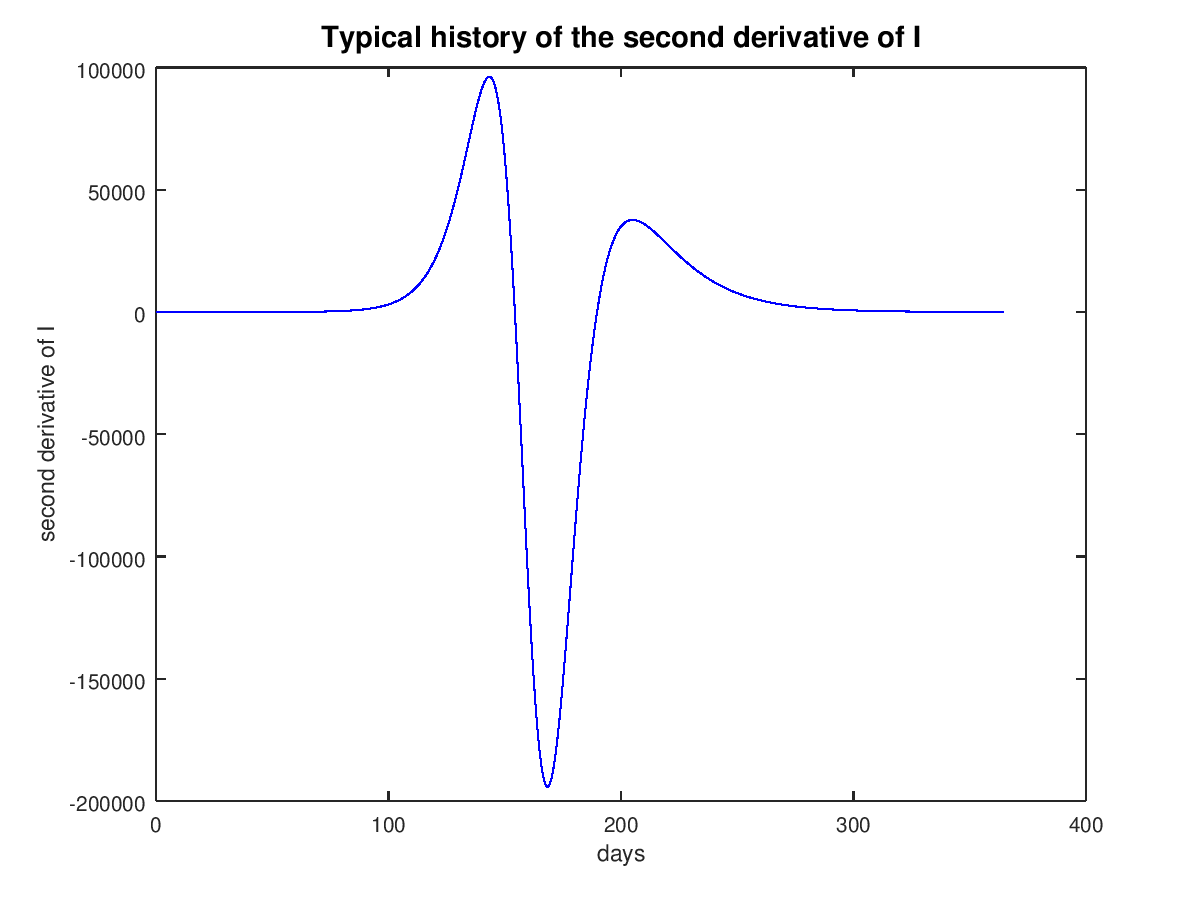}
\caption{\label{figusa4} History of the second derivative of $I$}
\end {center}
\end{figure}

The result of a dynamical 30 days lockdown for the UK is shown in fig. \ref{figuk2}, where we found $t_0 = 117$ ($\kappa = 0,2$).

Data from China and South Korea suggests that the group of infected people with an age
of 70 or more is of magnitude 10\%. This group has a significant higher mortality rate
than the rest of the infected people.
Thus we can presume that $\alpha$=10\% of $I$ must be especially sheltered and
possibly medicated very intensively as a high-risk group.

\begin{figure}[h]
\begin {center}
\includegraphics[width=10cm]{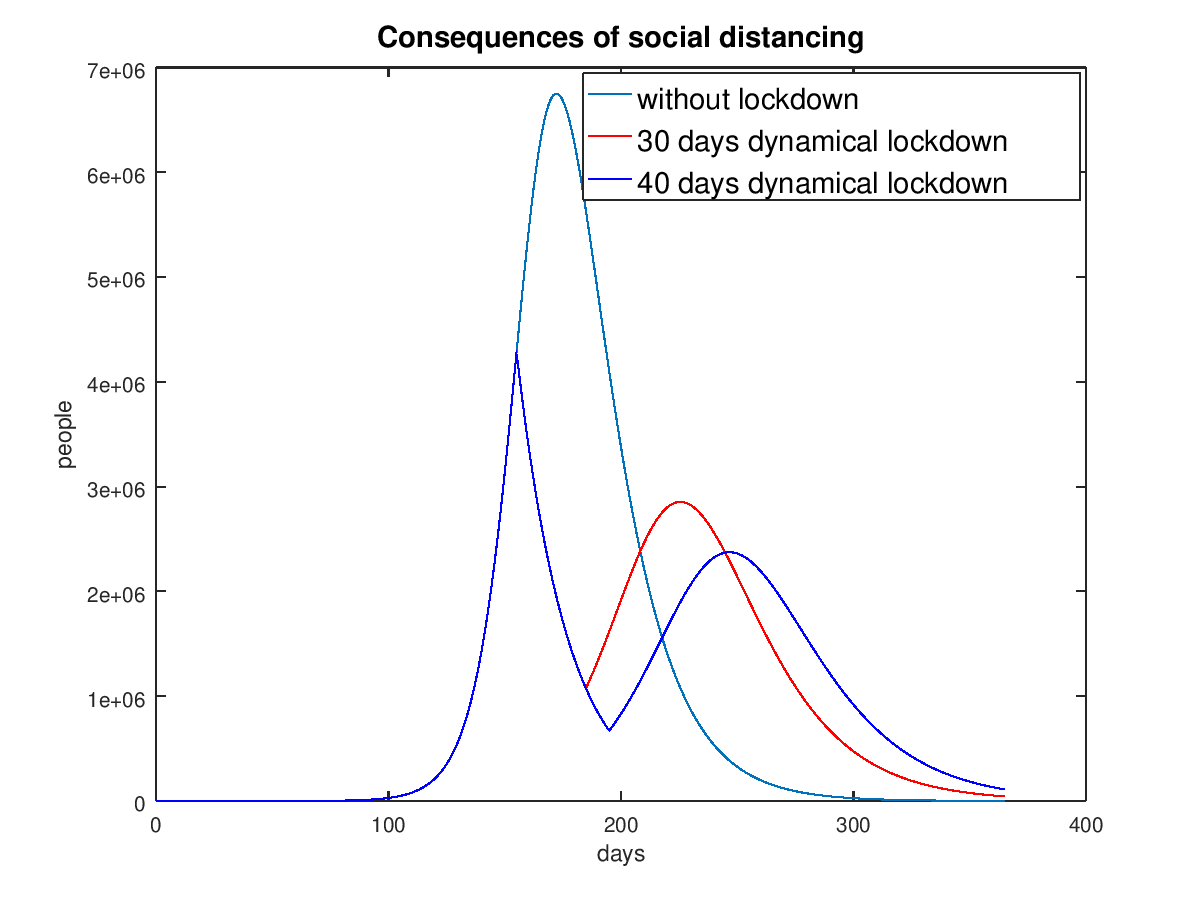}
\caption{\label{figusa5} USA history of the infected people of high-risk groups
depending on a dynamical lockdown}
\end {center}
\end{figure}

Fig. \ref{figusa5} shows the USA time history of the above defined high-risk group with a
dynamical lockdown with $\kappa = 0,2$ compared to regime without social distancing.
The maximum number of infected people decreases from approximately 6,7 millions of people to 4,2 millions
in the case of the lockdown (30 days lockdown).

This result proves the usefulness of a lockdown or a strict social distancing during
an epidemic disease.
We observe a flattening of the infection curve as requested by politicians and health professionals.
With a strict social distancing for a limited time
one can save time to find vaccines and time to improve the
possibilities to help high-risk people in hospitals.

\begin{figure}[h]
\begin {center}
\includegraphics[width=10cm]{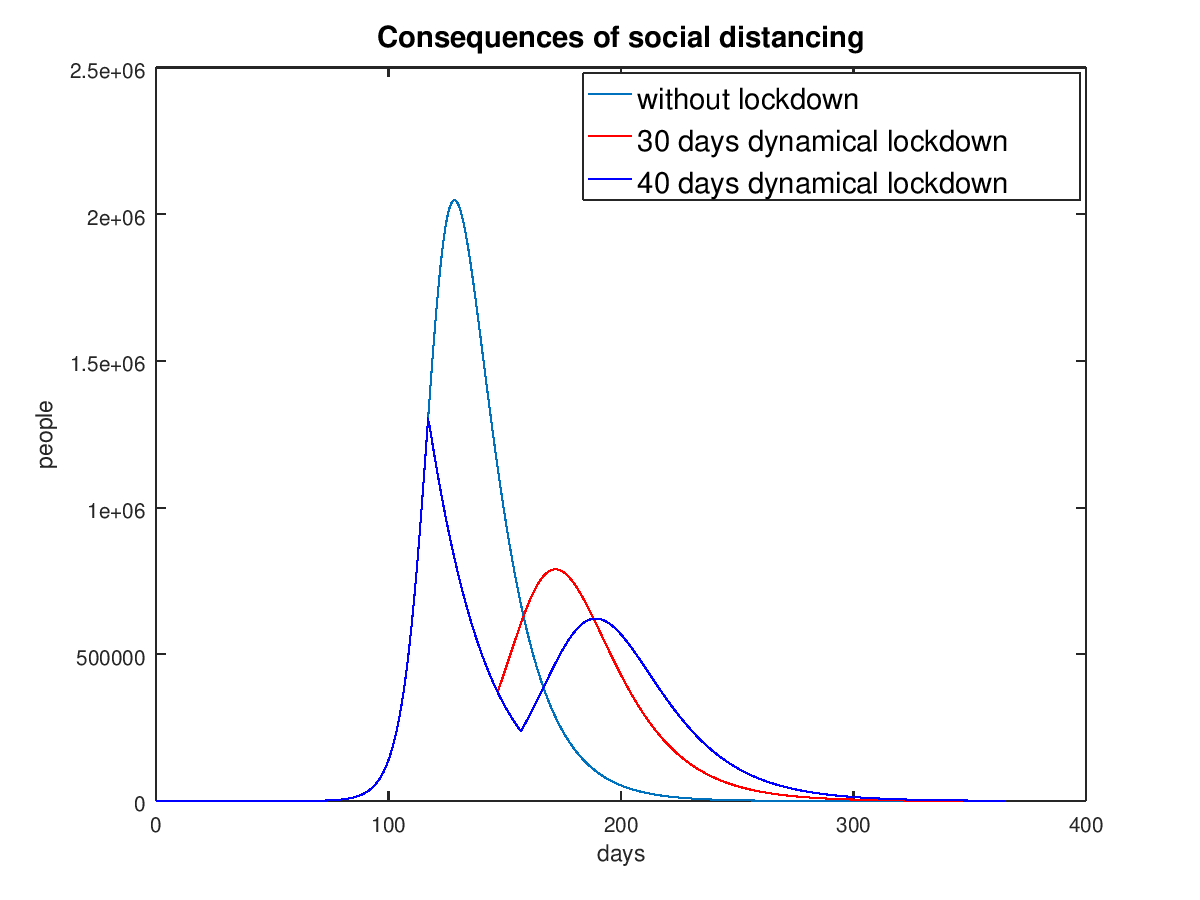}
\caption{\label{figuk3} UK history of the infected people of high-risk groups
depending on a dynamical lockdown}
\end {center}
\end{figure}

To see the influence of a social distancing we look at the UK situation without
a lockdown and a dynamical lockdown of 30 days with fig. \ref{figuk3} ($\kappa = 0,2$)
for the 10\% high-risk people.

The computations with the SIR model show, that the social distancing with a lockdown
will only be successful with a start behind the time greater or equal to $t_0$, found by the evaluation of the
second derivative of $I$ (formula \eqref{t0f}). If the lockdown is started at a time less then $t_0$ the
effect of such a social distancing is not significant.

\section{Closing remarks}
If we write \eqref{eq2} or \eqref{eq5} resp. in the form
\[
  \frac{d I}{dt}  =  (\kappa\beta \frac{S}{N}  - \gamma )I  \]
we realize that the number of infected people decreases if
\begin{equation}\label{eq7}
  \kappa\beta \frac{S}{N}  - \gamma < 0 \Longleftrightarrow S < N \frac{\gamma}{\kappa\beta}
\end{equation}
is complied.
The relation  \eqref{eq7} shows that there are two possibilities for the rise
of infected people to be inverted and the medical burden to be reduced.
\begin{itemize}
\item[a)] The reduction of the stock of the species $S$.
This can be obtained by immunization or vaccination.
Another possibility is the isolation of high-risk people (70 years and older).
Positive tests for antibodies reduce the stock of susceptible persons.
\item[b)] A second possibility is the reduction of the infection
rate $\kappa\beta$. This can be achieved by
strict lockdowns, social distancing at appropriate times, or rigid sanitarian moves.
\end{itemize}
With respect to point a) it is important to note, that a lot of
possible positive precautions by physicians and politicians can not be cover
by mathematical models like the SIR one.
Also the infected people are not distributed in the same way
at all locations of a country. It is also possible and necessary to concentrate the
modeling to hot spots like New York in the USA, Madrid in Spain or Bavaria in Germany
to get a higher resolution of the pandemic behavior.

The results are pessimistic in total with respect to a successful fight against
the COVID-19-virus. 
Hopefully the reality is a bit more merciful than the mathematical model.
But we rather err on the pessimistic side and be surprised by more benign developments.

Note again that the parameters $\beta$ and $\kappa$ are guessed very roughly.
Also, the percentage $\alpha$ of the group of high-risk people is possibly overestimated. Depending
on the capabilities and performance of the health system of the respective countries, those parameters
may look different.
The interpretation of $\kappa$ as a random variable is thinkable, too.

At the end all precautions (for example social distancing, isolation of high-risk people) lead to a prolongation of the pandemic period with respect
to the awaited and necessary herd immunity. But the decrease of the
peak of the curve of infected people generates time for the improvement
of the health systems and heights the possibilities to save life.


\begin{thebibliography}{9}
\bibitem{kak} W.O. Kermack and A.G. McKendrick, A contribution
to the mathematical theory of epidemics. Proc. R. Soc. London A 115(1927)700.
\bibitem{ecdc} Bulletins of the European Centre for Disease Prevention and Control (https://www.ecdc.europa.eu/en/geographical-distribution-2019-ncov-cases) 2020.
\bibitem{jhu} Bulletins of the John Hopkins University of world-wide Corona data (https://www.jhu.edu) 2020.
\bibitem{gb} G. B\"arwolff, Numerics for engineers, physicists and computer scientists (3rd ed., in German).
Springer-Spektrum 2020.
\bibitem{gb1} G. B\"arwolff, A Contribution to the Mathematical Modeling of the Corona/COVID-19 Pandemic.
medRxiv.preprint 2020, doi: https://doi.org/10.1101/2020.04.01.20050229.
\bibitem{tt} Toshihisa Tomie, Understandig the present status and forcasting of COVID-19 in Wuhan.
medRxiv.preprint 2020.
\end{thebibliography}
\end{document}